\documentclass[pra,twocolumn,showpacs,floatfix,preprintnumbers,nofootinbib]{revtex4}
\usepackage{amsmath,amssymb,bbm,bbold}
\usepackage{graphicx}
\usepackage{color}
\usepackage[normalem]{ulem}

\newcommand{\cev}[1]{\reflectbox{\ensuremath{\vec{\reflectbox{\ensuremath{#1}}}}}}

\begin{document}

\title{Quantum phase-space representation for curved configuration spaces}

\author{Clemens Gneiting}
\affiliation{Institute of Physics, Albert-Ludwigs University of Freiburg, Hermann-Herder-Stra{\ss}e 3, 79104 Freiburg, Germany}
\author{Timo Fischer}
\affiliation{University of Duisburg-Essen, Lotharstra{\ss}e 1-21, 47057 Duisburg, Germany}
\author{Klaus Hornberger}
\affiliation{University of Duisburg-Essen, Lotharstra{\ss}e 1-21, 47057 Duisburg, Germany}

%\date{\today}

\begin{abstract}
We extend the Wigner-Weyl-Moyal phase-space formulation of quantum mechanics to general curved configuration spaces. The underlying phase space is based on the chosen coordinates of the manifold and their canonically conjugate momenta. The resulting Wigner function displays the axioms of a quasiprobability distribution, and any Weyl-ordered operator gets associated with the corresponding phase-space function, even in the absence of continuous symmetries. The corresponding quantum Liouville equation reduces to the classical curved space Liouville equation in the semiclassical limit. We demonstrate the formalism for a point particle moving on two-dimensional manifolds, such as a paraboloid or the surface of a sphere. The latter clarifies the treatment of compact coordinate spaces, as well as the relation of the presented phase-space representation to symmetry groups of the configuration space.
\end{abstract}


%02.20.Qs     General properties, structure, and representation of Lie groups
%03.65.-w     Quantum mechanics
%03.65.Ca     Formalism
%05.30.-d     Quantum statistical mechanics
%33.80.-b     Photon interactions with molecules (see also 42.50.-p Quantum optics)
%37.90.+j     Other topics in mechanical control of atoms, molecules, and ions (restricted to new topics in section 37)
%42.50.-p     Quantum optics (for lasers, see 42.55.-f and 42.60.-v; see also 42.65.-k Nonlinear optics; 03.65.-w Quantum mechanics)
%42.50.Ct     Quantum description of interaction of light and matter; related experiments
%03.65.Sq Semiclassical theories in quantum mechanics,

\preprint{\textsf{published in Phys.~Rev.~A~{88}, 062117 (2013)}}

\maketitle

\section{Introduction}

The successful treatment of a physical problem often rests on the right choice of formalism, and conceptual progress is hard if one uses an inappropriate language. This is particularly evident in quantum theory, where diverging, yet equivalent formalisms come along with inherent advantages and disadvantages. For good reasons, the hydrogen atom is usually quantized by means of Schr\"odinger's wave equation, while fields are naturally described by Heisenberg operators. Feynman's path integral formulation, on the other hand, is often preferred in semiclassical contexts, and vast perturbative expansions are best handled diagrammatically. It is clearly useful to command a variety of alternative methods for treating a given physical system.

Wigner's prescription \cite{Wigner1932a} to transform the statistical operator into a distribution on classical phase space initiated yet another equivalent description of quantum mechanics. No other formalism is as powerful in highlighting quantum features while providing us with intuition in classical terms. Moyal completed it \cite{Moyal1949a} by discovering that Wigner's one-to-one mapping between quantum operators and functions on a classical phase space reflects Weyl's correspondence rule \cite{Weyl1928a}. This picture has found numerous applications in many areas of physics, ranging from solid-state physics to quantum optics \cite{Frensley1987a, Bastiaans1979a, Leonhardt1995a, Schleich2001a, Lutterbach1997a, Bertet2002a}, from particle interferometry to molecular physics \cite{Kurtsiefer1997a, Lee1980a, Wang1998a}, and from semiclassics to the foundations of quantum mechanics \cite{Almeida1990a, Dittrich2006a, Braunstein1998a, Banaszek1998a, Spekkens2008a,Mari2011a}; it has even been adapted to quantum field theory \cite{Vasak1987a, Bialynicki1991a, Zhuang1996a, Hebenstreit2011a}.

In this article we generalize the standard Cartesian Wigner-Weyl-Moyal formalism to general curved configuration spaces. It applies to highly symmetric surfaces such as spheres or paraboloids, but also to arbitrary manifolds without any symmetry constraints.  
All the characteristic features of the Cartesian case carry over. Along with the interpretation of the Wigner function as a quasiprobability distribution, these include the replacement of operator traces by phase-space integrals and a meaningful semiclassical limit. As we demonstrate, the phase-space coordinates must be constructed from canonically conjugate operator pairs. The associated translation operators are used to define the Stratonovich-Weyl operator kernels, which lie at the heart of the Wigner-Weyl-Moyal formalism \cite{Stratonovich1956a,Stratonovich1957a}. In this sense, the formalism is consistently based on the group of translations, independent of whether these reflect an isometry of the manifold or whether competing dynamical symmetry groups are present.

Our phase space is based on coordinates $x^i$ of the Riemannian manifold and their canonically conjugate momenta $p_i$. It does provide by construction an unambiguous partitioning into mutually independent pairs of phase space variables, reflected by the classical canonical Poisson brackets $\{x^i,p_j\}=\delta^i_j$ and $\{x^i,x^j\}=\{p_i,p_j\}=0$. This guarantees that the resulting phase space representation behaves in correspondence to its classical counterpart, as in the Cartesian case. We then find that any Weyl-ordered observable is mapped to its equivalent function on phase space, that integrating out phase space coordinates yields the marginal probability distribution, and that the motion of the Wigner function is described by a quantum Liouville equation, which turns into its classical equivalent as $\hbar \rightarrow 0$.

Various attempts to generalize the Cartesian Wigner function considered highly symmetric manifolds, in particular homogeneously curved spaces such as spheres \cite{Agarwal1981a, Varilly1989a, Figueroa1990a, Dowling1994a, Brif1998a, Brif1999a, Nasyrov1999a, Mukunda2004a, Mukunda2005a, Klimov2008a}. These approaches focused on the underlying symmetry groups to construct a phase space. But the generators $\hat{J}_i$ of a Lie group obey generally non-canonical commutation relations, $[\hat{J}_r,\hat{J}_s] = {\rm i} C_{r s}^t \hat{J}_t$, involving the structure constants $C_{r s}^t$ of the group. If taken as momenta, these generators interfere detrimentally among each other, unless one deals with the translation group. None of the resulting phase space representations preserves all of the above mentioned essential features of the Cartesian formalism. Other approaches are based on a mapping to the Cartesian case \cite{Pierre1969a, Nienhuis1970a, Nienhuis1970b, Bondar2012a}, on eigenfunctions of the Laplace-Beltrami operator in the case of hyperboloids \cite{Alonso2002a}, or on a covariant, field-theoretic extension to a generalized density matrix \cite{Winter1985a}. All the relevant properties of the standard Wigner function have so far been demonstrated only for the one-dimensional case of a single angle-angular momentum pair, which is a curvature-free problem \cite{Mukunda1978a, Mukunda1979a, Leonhardt1995b, Leonhardt1996a, Rigas2011a}, and for the orientation state of a rigid body \cite{Fischer2013a}, which is an instance of the general theory presented in this article.

Many physical systems exhibit compact configuration spaces. Simple examples are the one-dimensional motion of a point particle constrained to a circle, or the two-dimensional motion of a particle confined to the surface of a sphere. Underlying symmetry groups are then compact, with discrete, finite-dimensional irreducible representations. Likewise, the canonically conjugate momentum operators generating coordinate translations then exhibit discrete spectra. This is a consequence of the compactness and is not directly related to the curvature of the configuration space. We demonstrate how the presented phase space formalism is applied to configuration spaces implying discrete momentum spectra.

This article is structured as follows: In Section \ref{Section:Cartesian_Wigner_function} we recapitulate the Wigner-Weyl-Moyal representation for the Cartesian case and briefly demonstrate how to construct it from Stratonovich-Weyl operator kernels. Quantum mechanics in curved configuration spaces is introduced in Section \ref{Section:Quantum_mechanics_on_curved_configuration_spaces}, and we provide in Section \ref{Section:Curved_space_Wigner_function} the corresponding phase space description. In Section \ref{Section:Extended_two-dimensional_curved_spaces} we illustrate the presented Wigner-Weyl-Moyal formalism by means of two-dimensional curved surfaces embedded in three-dimensional Euclidean space. Section \ref{Section:Free_particle_on_a_sphere} then elaborates the formalism for compact configuration spaces, demonstrating it for a point particle constrained to the surface of a sphere. Finally, we present our conclusions in Section \ref{Section:Conclusions}.

\section{Cartesian Wigner function} \label{Section:Cartesian_Wigner_function}

It is instructive to briefly recapitulate the Wigner-Weyl formalism for a single point particle in Euclidean space, parametrized by Cartesian coordinates \cite{Schleich2001a}. The Wigner function of a quantum state $\hat{\rho}$ in a single space dimension then assumes the well-known form
\begin{equation} \label{CartesianWignerFunction}
W(x,p) = \frac{1}{2 \pi \hbar} \int_{-\infty}^{\infty} \mathrm{d}x' \mathrm{e}^{{\rm i} p x'/\hbar} \langle x-x'/2|\hat{\rho}|x+x'/2\rangle,
\end{equation}
where the phase space variables $x$ and $p$ are based on the position  and momentum operator $\hat{x}$ and $\hat{p}$, respectively, with $[\hat{x},\hat{p}]={\rm i} \hbar$. Similarly, one can write the Wigner function in momentum representation as
\begin{equation} \label{Cartesian_Wigner_function_momentum}
W(x,p) = \frac{1}{2 \pi \hbar} \int_{-\infty}^{\infty} \mathrm{d}p' \mathrm{e}^{-{\rm i} p' x/\hbar} \langle p-p'/2|\hat{\rho}|p+p'/2\rangle.
\end{equation}
The Wigner function (\ref{CartesianWignerFunction}) can easily be generalized to the case of $N$ point particles in $d$ space dimensions, but for the sake of clarity we confine the discussion here to a single degree of freedom.

\subsection{The Wigner function as a quasiprobability distribution}

It is straightforward to verify that the Wigner function (\ref{CartesianWignerFunction}) is real, $W(x,p) \in \mathbb{R} \hspace{2mm} \forall \, x,p$, and that it satisfies the {\em axioms of a quasi-probability distribution}:
\begin{subequations} \label{Axioms_quasi-probability_distribution}
\begin{align}
\int_{-\infty}^{\infty}\mathrm{d}x \int_{-\infty}^{\infty}\mathrm{d}p \, W(x,p) & = 1, \label{NormalizationAxiom} \\
\int_{-\infty}^{\infty}\mathrm{d}p \, W(x,p) & = \langle x|\hat{\rho}|x\rangle, \label{PositionMarginalAxiom} \\
\int_{-\infty}^{\infty}\mathrm{d}x \, W(x,p) & = \langle p|\hat{\rho}|p\rangle. \label{MomentumMarginalAxiom}
\end{align}
\end{subequations}
Equation (\ref{NormalizationAxiom}) reflects normalization, and Eqs.~(\ref{PositionMarginalAxiom}) and (\ref{MomentumMarginalAxiom}) describe the marginal properties, i.e.~the possibility to infer the probability distribution of a phase space variable by integrating out the respective conjugate variable.

By virtue of Eq.~(\ref{Axioms_quasi-probability_distribution}) the Wigner function behaves analogously to a classical probability distribution, which is one of its most outstanding features and sets it apart from other phase space representations, such as the Glauber $P$-function and the Sudarshan $Q$-function \cite{Walls2008a}. The substantial difference to a genuine, classical probability distribution arises when the Wigner function takes negative values. If such negativities occur, it can be taken as a signature of genuine quantum features. A prominent example is the coherent superposition of two spatially distinct wave packets, where their capability to interfere is captured by a partly negative interference term in the Wigner function.

\subsection{Weyl correspondence}

The phase-space description not only yields an appealing way to represent quantum states, but it provides us with a formulation of quantum mechanics which is completely equivalent to Hilbert space quantum mechanics; that is, all relevant objects and operations in quantum theory, such as states, observables, expectation values, time evolution etc., can be represented in phase space. In the following we briefly recapitulate how the Wigner-Weyl-Moyal phase space formalism can be constructed using Stratonovich-Weyl operator kernels.

One starts by defining displacement operators
\begin{equation} \label{CartesianDisplacementOperators}
\hat{D}(x,p) = \mathrm{e}^{-{\rm i} x \hat{p}/\hbar} \mathrm{e}^{{\rm i} p \hat{x}/\hbar}.
\end{equation}
These subsequently shift a quantum state by $p$ in momentum and translate it by a distance $x$. In that sense the displacement operators are composed of translations. Note that the displacement operators are often defined symmetrically as $\exp[{\rm i} (p \hat{x}-x \hat{p})/\hbar]$. Applying the Baker-Campbell-Hausdorff formula then yields Eq.~(\ref{CartesianDisplacementOperators}) with an additional, but irrelevant phase factor $\exp[(-{\rm i} x p/(2 \hbar)]$. While such a symmetric definition may appear appealing, it turns out that it cannot be easily generalized to situations where the commutators of the coordinates and their conjugate momenta are not $c$-numbers. This is already the case, e.g., for a single angle variable and its angular momentum \cite{Rigas2011a} (see also \cite{Mukunda2005a}).

Using the displacement operators (\ref{CartesianDisplacementOperators}), one introduces the Stratonovich-Weyl operator kernel \cite{Varilly1989a}
\begin{equation} \label{CartesianStratonovichWeylOperatorKernel}
\hat{\Delta}(x,p) = \hat{D}(x,p)\hat{\Delta}(0,0)\hat{D}^{\dagger}(x,p),
\end{equation}
with the undisplaced kernel chosen as
\begin{align} \label{CartesianUndisplacedStratonovichWeylOperatorKernel}
\hat{\Delta}(0,0) &= \frac{1}{2 \pi \hbar} \int \mathrm{d}p' \int \mathrm{d}x' \hat{D}(x',p') \mathrm{e}^{{\rm i} x' p'/2 \hbar} \\
&= \int \mathrm{d}x' |x'/2 \rangle \langle -x'/2|. \nonumber
\end{align}
Note that we could have omitted the phase factor in Eq.~(\ref{CartesianUndisplacedStratonovichWeylOperatorKernel}) if we had chosen the symmetric definition of the displacement operators. The operator $\hat{\Delta}(x,p)$ can be interpreted as effecting a displaced parity operation, since $\hat{\Delta}(0,0)$ is proportional to the parity operator $\hat{P}=\int \mathrm{d}x |x\rangle \langle -x|$, $\hat{\Delta}(0,0)=2 \hat{P}$. With Eq.~(\ref{CartesianUndisplacedStratonovichWeylOperatorKernel}) one gets
\begin{equation}
\hat{\Delta}(x,p) = \int \mathrm{d}x' {\rm e}^{{\rm i} p x'/\hbar} |x+x'/2 \rangle \langle x-x'/2|.
\end{equation}

The Stratonovich-Weyl operator kernel (\ref{CartesianStratonovichWeylOperatorKernel}) is the central tool to construct the complete phase space formalism. In particular, it constitutes a basis in operator space, as reflected by the completeness relation
\begin{equation}
{\rm tr}[\hat{\Delta}(x,p) \hat{\Delta}(\tilde{x},\tilde{p})] = 2 \pi \hbar \, \delta(x-\tilde{x}) \delta(p-\tilde{p}).
\end{equation}
Below we will see how it can be generalized to curved configuration spaces.

The phase space representatives of arbitrary Hilbert space operators $\hat{A}$, i.e.~the Weyl symbols $W_{\hat{A}}$, can now be obtained as
\begin{subequations} \label{Cartesian_Weyl_symbols}
\begin{align}
W_{\hat{A}}(x,p) =& \mathrm{tr}[\hat{A} \hat{\Delta}(x,p)] \\
 =& \int \mathrm{d}x' \mathrm{e}^{{\rm i} p x'/\hbar} \langle x-x'/2|\hat{A}|x+x'/2\rangle.
\end{align}
\end{subequations}
In the case of Hermitian operators $\hat{A}$ the Weyl symbols are real. Equation (\ref{Cartesian_Weyl_symbols}) establishes a one-to-one mapping from Hilbert space to phase space, where the inverse is given by
\begin{equation} \label{CartesianOperatorPhaseSpaceExpansion}
\hat{A} = \frac{1}{2 \pi \hbar} \int \mathrm{d}x \int \mathrm{d}p W_{\hat{A}}(x,p) \hat{\Delta}(x,p).
\end{equation}
Note that the Wigner function (\ref{CartesianWignerFunction}) is the Weyl symbol of the density operator $\hat{\rho}$, multiplied by $1/(2 \pi \hbar)$. This prefactor is introduced to normalize the Wigner function, see Eq.~(\ref{NormalizationAxiom}). Moreover, expectation values $\langle \hat{A} \rangle = {\rm tr}[\hat{\rho} \hat{A}]$ then follow from the phase-space integral
\begin{equation}
\langle \hat{A} \rangle = \int {\rm d}x \int {\rm d}p \, W_{\hat{A}}(x,p) W(x,p),
\end{equation}
in full analogy to a classical description. In the case of two arbitrary operators, ${\rm tr}[\hat{A} \hat{B}]$, this is referred to as the tracing condition.

The Weyl symbol of the product of two operators is obtained from their individual Weyl symbols via the star product
\begin{align}
W_{\hat{A} \hat{B}}(x,p) =& (W_{\hat{A}} \star W_{\hat{B}})(x,p) \\
 =& \int \frac{{\rm d}x_1 {\rm d}p_1}{\pi \hbar} \int \frac{{\rm d}x_2 {\rm d}p_2}{\pi \hbar} {\rm e}^{2 {\rm i} (x_1 p_2-x_2 p_1)/\hbar} \nonumber \\
 & \times W_{\hat{A}}(x+x_1, p+p_1) W_{\hat{B}}(x+x_2, p+p_2). \nonumber
\end{align}
Equivalent ways to express the star product are
\begin{subequations} \label{Cartesian_differential_star_product}
\begin{align}
(W_{\hat{A}} \star& W_{\hat{B}})(x,p) \nonumber \\
&= W_{\hat{A}}\left(x+\frac{{\rm i} \hbar \vec{\partial}_p}{2}, p-\frac{{\rm i} \hbar \vec{\partial}_x}{2} \right) W_{\hat{B}}(x,p) \\
&= W_{\hat{A}}(x,p) W_{\hat{B}}\left(x-\frac{{\rm i} \hbar \cev{\partial}_p}{2}, p+\frac{{\rm i} \hbar \cev{\partial}_x}{2} \right).
\end{align}
\end{subequations}
The arrows on top of the differential operators indicate that they act only on the respective other Weyl symbol. The star product prescribes how to obtain the Weyl symbols of arbitrary operator expressions from the Weyl symbols of their elementary constituents. The expressions (\ref{Cartesian_differential_star_product}) are useful to derive quantum Liouville equations.

The most elementary Weyl symbols are the phase-space representations of the position and the momentum operator,
\begin{equation}
W_{\hat{x}}(x,p) = x \hspace{2mm} \text{and} \hspace{2mm} W_{\hat{p}}(x,p) = p.
\end{equation}
These Weyl symbols are identical with the corresponding classical phase space functions, which again manifests the close analogy of the Wigner-Weyl-Moyal phase space formalism with the classical phase space description. Importantly, this also holds for arbitrary Weyl-ordered moments,
\begin{equation} \label{Weyl_ordered_xp}
W_{\{\hat{p}^k, \hat{x}^{\ell}\}_{\text{W}}}(x,p) = p^k x^{\ell}.
\end{equation}
The Weyl ordering is defined as
\begin{equation} \label{Weyl_ordering}
\{\hat{p}^k, \hat{x}^{\ell}\}_{\rm W} = 2^{-k} \sum_{j=0}^{k} \left( \begin{array}{c} k \\ j \end{array} \right) \hat{p}^{k-j} \hat{x}^{\ell} \hat{p}^{j};
\end{equation}
in the Cartesian case it is equivalent to the symmetric ordering. Below, we will see that the relation (\ref{Weyl_ordered_xp}) can be extended to curved configuration spaces. It is now easy to see that the Weyl symbol of a Hamiltonian $\hat{H} = \hat{p}^2/2 m+V(\hat{x})$ is given by
\begin{equation}
W_{\hat{H}}(x,p) = \frac{p^2}{2 m} + V(x),
\end{equation}
which coincides with the classical Hamiltonian.

\subsection{The quantum Liouville equation}

So far we have discussed kinematic aspects. Let us now consider the dynamics of a quantum point particle in terms of its phase-space description. The von Neumann equation, ${\rm i} \hbar \partial_t \hat{\rho} = [\hat{H},\hat{\rho}]$, then translates into its phase-space version,
\begin{equation}
\frac{\partial W(x,p)}{\partial t} = -\frac{{\rm i}}{\hbar} [W_{\hat{H}} \star W(x,p)-W\star W_{\hat{H}}(x,p)].
\end{equation}
For a particle of mass $m$ described by a Hamiltonian of the form $\hat{H} = \hat{p}^2/2 m+V(\hat{x})$ and using Eq.~(\ref{Cartesian_differential_star_product}) one obtains the quantum Liouville equation
\begin{align} \label{CartesianQuantumLiouvilleEquation}
\Bigg(\partial_t &+\frac{p}{m} \frac{\partial}{\partial x}-\frac{\mathrm{d} V(x)}{\mathrm{d} x} \frac{\partial}{\partial p}\Bigg) W(x,p) \nonumber \\
= &\sum_{\ell=1}^{\infty} \frac{(-1)^{\ell} (\hbar/2)^{2 \ell}}{(2 \ell+1)!} \frac{\mathrm{d}^{2 \ell+1} V(x)}{\mathrm{d}x^{2 \ell+1}} \frac{\partial^{2 \ell+1}}{\partial p^{2 \ell+1}} W(x,p).
\end{align}

To leading order in $\hbar$ the right hand side vanishes so that one obtains the classical Liouville equation for the corresponding Hamilton function $H(x,p)=p^2/2m+V(x)$. The right hand side thus generates quantum corrections to the classical time evolution. It vanishes if the potential is at most harmonic. The dynamics generated by the quantum Liouville equation (\ref{CartesianQuantumLiouvilleEquation}) is then identical with the classical time evolution, a feature often helpful in practical applications. In particular, the time evolution of a free particle merely shears the Wigner function.

\section{Quantum mechanics on curved configuration spaces} \label{Section:Quantum_mechanics_on_curved_configuration_spaces}

We now proceed to the general situation of curved configuration spaces. We start by clarifying important kinematic and dynamic aspects of quantum mechanics in curved spaces.

\subsection{Hilbert space of a curved manifold}

Throughout this section, we consider an $n$-dimensional Riemannian manifold with metric tensor $g_{i j}(x)$ and associated determinant
\begin{equation}
g(x) = \det g_{i j}(x).
\end{equation}
For simplicity, we assume that we have a global parametrization of the manifold with coordinates $x^i, i=1,\dots, n$. Moreover, let us assume that the configuration space is unbounded, so that the coordinates $x^i$ are from an unbounded interval supporting canonically conjugate momenta with continuous spectra. In Sec.~\ref{Section:Free_particle_on_a_sphere}, we also discuss examples where the configuration space is compact.

To ease notation, we abbreviate $\int \mathrm{d}x f(x) \equiv \int \mathrm{d}x^1\dots \mathrm{d}x^n f(x^1,\dots,x^n)$, $\delta(x-x') \equiv \delta(x^1-x'^1)\dots \delta(x^n-x'^n)$, and we use the Einstein sum convention. The identity operator expressed in terms of the coordinate basis then reads as \cite{DeWitt1952a, DeWitt1957a}
\begin{equation} \label{Coordinate_basis_curved}
\mathbb{1} = \int \mathrm{d}x \sqrt{g(x)} \, |x\rangle \langle x|. 
\end{equation}
Note that due to the metric determinant it is in general not possible to decompose Eq.~(\ref{Coordinate_basis_curved}) into a tensor product of single-coordinate Hilbert spaces. The coordinate eigenstates exhibit the orthogonality relation
\begin{equation} \label{Coordinate_orthogonality_curved}
\langle x|x' \rangle = \frac{1}{\sqrt{g(x)}} \delta(x-x'),
\end{equation}
and the coordinate operators $\hat{x}^i$ satisfy
\begin{equation}
\hat{x}^i |x\rangle = x^i |x\rangle.
\end{equation}

\subsection{Conjugate momentum basis}

We now seek an equivalent Hilbert space representation in terms of the canonically conjugate momenta. Following DeWitt \cite{DeWitt1952a, DeWitt1957a}, one obtains the conjugate momentum operators $\hat{p}_i$ from the requirement that they shall satisfy the canonical commutation relations $[\hat{x}^i,\hat{p}_i]={\rm i} \hbar \delta_{\,j}^i$ and $[\hat{p}_i,\hat{p}_j]=0$. This yields the quantization rule
\begin{equation} \label{CurvedSpaceMomentumQuantizationRule}
p_i \rightarrow \hat{p}_i = \frac{\hbar}{{\rm i}} \left(\frac{\partial}{\partial x^i}+\frac{1}{2} \Gamma_{j i}^j(x) \right),
\end{equation}
where the curvature is reflected by the contracted Christoffel symbol $\Gamma_{j i}^j(x)$. 

The Christoffel symbols of a Riemannian manifold are defined by the contravariant and covariant components of the metric tensor,
\begin{equation}
\Gamma^k_{i j} = \frac{1}{2} g^{k l} \left(\frac{\partial g_{j l}}{\partial x^i}+\frac{\partial g_{i l}}{\partial x^j}-\frac{\partial g_{i j}}{\partial x^l} \right).
\end{equation}
It follows that the contracted Christoffel symbol is given by the simple expression
\begin{equation}
\Gamma_{j i}^j(x) = \frac{1}{2} \frac{\partial \ln g(x)}{\partial x^i}.
\end{equation}

Compared to the Cartesian quantization rule, $p_i \rightarrow \hat{p}_i= -{\rm i} \hbar \, \partial/\partial x^i$, the ordinary partial derivative is thus replaced by the derivative (\ref{CurvedSpaceMomentumQuantizationRule}) (resembling a covariant derivative up to the factor $1/2$). The momentum eigenstates in coordinate representation now follow from the eigenvalue equations ($i=1,\dots,n$)
\begin{equation}
\frac{\hbar}{{\rm i}} \left(\frac{\partial}{\partial x^i}+\frac{1}{2} \Gamma_{j i}^j(x) \right) \langle x|p \rangle = p_i \langle x|p \rangle,
\end{equation}
which are solved by
\begin{equation} \label{Momentum_eigenstates_curved}
\langle x|p \rangle = \frac{\mathrm{e}^{{\rm i} p_i x^i/\hbar}}{(2 \pi \hbar)^{n/2} \sqrt[4]{g(x)}}.
\end{equation}
Similar to the Cartesian case, the momentum eigenstates span the Hilbert space, and the unity operator expressed in this basis reads
\begin{equation} \label{Momentum_basis_curved}
\mathbb{1} = \int \mathrm{d}p \, |p\rangle \langle p|.
\end{equation}

Below we see that the canonically conjugate momentum operators $\hat{p}_i$ and their eigenstates $|p\rangle$ play a central role for defining the curved space Wigner function. We remark that in compact coordinate spaces (such as a circle or a sphere) the commutators between conjugate operator pairs are necessarily operator valued. The quantization rule (\ref{CurvedSpaceMomentumQuantizationRule}) remains valid in this case, while the compactness is then reflected by discrete momentum spectra.

\subsection{Quantum Hamiltonian}

In order to formulate the quantum dynamics we need to set up the quantum Hamiltonian. Here one should keep in mind that there is no unique prescription to quantize the classical curved space Hamilton function
\begin{equation} \label{Classical_Hamiltonian_curved}
H(x,p) = \frac{1}{2 m} g^{i j}(x) p_i p_j + V(x).
\end{equation}
This is because different Hermitian operator orderings of the kinetic energy term result in different, inequivalent quantum corrections to the potential. Ultimately, the correct Hamiltonian can only be confirmed empirically. We stress that the issue of identifying the correct quantum Hamiltonian is not related to our task of finding a viable phase space description for curved configuration spaces, since we can assume the quantum Hamiltonian to be given.

DeWitt derived the quantum correction for the case of a specific operator ordering 
\begin{equation} \label{CurvedSpaceHamiltonian}
\hat{H} = \frac{1}{2 m} \hat{p}_i g^{i j}(\hat{x}) \hat{p}_j+\hbar^2 Q(\hat{x})+V(\hat{x}).
\end{equation}
It involves a so-called quantum potential $Q(x)$, which guarantees the covariance of the Hamiltonian and reads as \cite{DeWitt1952a, DeWitt1957a}
\begin{equation} \label{Curved_quantum_potential}
Q(x) = \frac{1}{4 m} g^{i j} \left[ \frac{\partial}{\partial x^j}\Gamma_{k i}^k-\Gamma^k_{i j} \Gamma^l_{l k}-\frac{1}{2} \Gamma^k_{k i} \Gamma^l_{l j} \right].
\end{equation}
For the sake of concreteness, we assume in the following that the quantum Hamiltonian takes the form (\ref{CurvedSpaceHamiltonian}); it is clear that other choices of operator ordering can be treated similarly. We emphasize that the occurrence of a quantum potential is a generic phenomenon in curved spaces. It must be taken into account to obtain the correct quantum dynamics. In Sec.~\ref{Section:Extended_two-dimensional_curved_spaces} we see how it arises in the case of a particle constrained to the surface of a sphere and a paraboloid.

Let us remark that similar ambiguities arise if one seeks to quantize the kinetic energy using the Laplace-Beltrami operator,
\begin{equation} \label{Laplace-Beltrami_operator}
\Delta f = \frac{1}{\sqrt{|g|}} \frac{\partial}{\partial x^i} \left( \sqrt{|g|} g^{i j} \frac{\partial}{\partial x^j} f \right).
\end{equation}
Here, the ambiguity arises in that one is free to add scalar curvature terms without affecting covariance of the kinetic energy \cite{Marinov1980a}. The Laplace-Beltrami operator (\ref{Laplace-Beltrami_operator}) is equivalent to DeWitt's choice (\ref{CurvedSpaceHamiltonian}), as shown in Ref.~\cite{Marinov1980a}. The expression of the kinetic energy in Eq.~(\ref{CurvedSpaceHamiltonian}) is favorable in order to determine its Weyl symbol.

\section{Curved-space Wigner function} \label{Section:Curved_space_Wigner_function}

In the following we derive the Wigner function for curved configuration spaces. This can be achieved following a similar line of argument as for the Cartesian Wigner function, introducing displacement operators and Stratonovich-Weyl quantizers. For this it is crucial that the phase-space representation is based on mutually commuting conjugate operator pairs.

\subsection{Stratonovich-Weyl operator kernel}

In the previous section we discussed the momentum operators canonically conjugate to the curved space coordinates. As a defining feature of such conjugate operator pairs, they mutually generate translations in their conjugate coordinate. In the general curved case the translation operators act according to
\begin{subequations}
\begin{align}
{\rm e}^{-{\rm i} x^i \hat{p}_i/\hbar} |x'\rangle &= \sqrt[4]{\frac{g(x'+x)}{g(x')}} |x'+x\rangle, \\
{\rm e}^{{\rm i} p_i \hat{x}^i/\hbar} |p'\rangle &= |p'+p\rangle,
\end{align}
\end{subequations}
as follows from the representations of unity (\ref{Coordinate_basis_curved}) and (\ref{Momentum_basis_curved}). In analogy to Eq.~(\ref{CartesianDisplacementOperators}), we now consider the displacement operators
\begin{equation} \label{Curved_Displacement_Operators}
\hat{D}^{(g)}(x,p) = {\rm e}^{-{\rm i} x^i \hat{p}_i/\hbar} {\rm e}^{{\rm i} p_i \hat{x}^i/\hbar}.
\end{equation}
Next, these are used to define the undisplaced operator kernel in analogy to Eq.~(\ref{CartesianUndisplacedStratonovichWeylOperatorKernel}),
\begin{align} \label{Curved_undisplaced_Stratonovich_Weyl_operator_kernel}
\hat{\Delta}^{(g)}(0,0) &= \frac{1}{(2 \pi \hbar)^n} \int {\rm d}p' \int {\rm d}x' \hat{D}^{(g)}(x',p') {\rm e}^{{\rm i} x'^i p'_i/2 \hbar} \nonumber \\
 &= \int {\rm d}x' \sqrt[4]{g(-x'/2) g(x'/2)} \, |x'/2\rangle \langle -x'/2|.
\end{align}
As in Eq.~(\ref{CartesianStratonovichWeylOperatorKernel}), we can finally introduce the Stratonovich-Weyl operator kernel by
\begin{equation} \label{Curved_Stratonovich_Weyl_quantizer_definition}
\hat{\Delta}^{(g)}(x,p) = \hat{D}^{(g)}(x,p) \hat{\Delta}^{(g)}(0,0) \hat{D}^{(g) \dagger}(x,p),
\end{equation}
which evaluates as
\begin{align} \label{Curved_Stratonovich_Weyl_quantizer_evaluated}
\hat{\Delta}^{(g)}(x,p) =& \int {\rm d}x' \sqrt[4]{g(x+x'/2) g(x-x'/2)} \nonumber \\
& \times {\rm e}^{{\rm i} p_i x'^i/\hbar} |x+x'/2\rangle \langle x-x'/2|.
\end{align}
A nontrivial metric determinant $g$ thus modifies the Stratonovich-Weyl operator kernel compared to the Cartesian one. The completeness  in operator space, on the other hand, remains untouched,
\begin{equation}
{\rm tr}[\hat{\Delta}^{(g)}(x,p) \hat{\Delta}^{(g)}(\tilde{x},\tilde{p})] = (2 \pi \hbar)^n \, \delta(x-\tilde{x}) \delta(p-\tilde{p}).
\end{equation}
As we will see, this ensures that the equivalence of the phase-space representation with Hilbert space quantum mechanics is maintained in the general curved case.

\subsection{Wigner function}

The Wigner function in a curved space is obtained from the Stratonovich-Weyl quantizer (\ref{Curved_Stratonovich_Weyl_quantizer_evaluated}) according to
\begin{align} \label{CurvedWignerFunction}
W^{(g)}(x,p) =& \frac{1}{(2 \pi \hbar)^n} {\rm tr}[\hat{\rho} \hat{\Delta}^{(g)}(x,p)] \nonumber \\
 =& \frac{1}{(2 \pi \hbar)^n} \int {\rm d}x' \sqrt[4]{g(x+x'/2) g(x-x'/2)} \nonumber \\
& \times {\rm e}^{{\rm i} p_i x'^i/\hbar} \langle x-x'/2|\hat{\rho}|x+x'/2\rangle.
\end{align}
It is the natural generalization of the Cartesian Wigner function (\ref{CartesianWignerFunction}). Formally, the momentum representation of the Wigner function remains unchanged (cf. Eq.~(\ref{Cartesian_Wigner_function_momentum})),
\begin{equation}
W^{(g)}(x,p) = \frac{1}{(2 \pi \hbar)^n} \int \mathrm{d}p' \mathrm{e}^{-{\rm i} p_i' x^i/\hbar} \langle p-p'/2|\hat{\rho}|p+p'/2\rangle,
\end{equation}
although one should keep in mind that the position representation of the momentum eigenstates is modified, see (\ref{Momentum_eigenstates_curved}).

One easily verifies that the curved-space Wigner function (\ref{CurvedWignerFunction}) maintains the axioms of a quasiprobability distribution,
\begin{subequations} \label{Curved_Wigner_marginals}
\begin{align}
\int {\rm d}x \int {\rm d}p \, W^{(g)}(x,p) &= 1, \\
\int {\rm d}p \, W^{(g)}(x,p) &= \sqrt{g(x)} \langle x|\hat{\rho}|x\rangle, \\
\int {\rm d}x \, W^{(g)}(x,p) &= \langle p|\hat{\rho}|p\rangle.
\end{align}
\end{subequations}

\subsection{Weyl symbols}

In analogy to the Cartesian case, the Weyl symbol of an arbitrary operator $\hat{A}$ is defined by
\begin{subequations} \label{Curved_Weyl_symbol}
\begin{align}
W_{\hat{A}}^{(g)}(x,p) = & {\rm tr}[\hat{A} \, \hat{\Delta}^{(g)}(x,p)] \\
= & \int {\rm d}x' \sqrt[4]{g(x+x'/2) g(x-x'/2)} \nonumber \\
 & \times {\rm e}^{{\rm i} p_i x'^i/\hbar} \langle x-x'/2|\hat{A}|x+x'/2\rangle,
\end{align}
\end{subequations}
and the inverse mapping is given by
\begin{equation} \label{Curved_Weyl_inverse}
\hat{A} = \frac{1}{(2 \pi \hbar)^n} \int {\rm d}x \int {\rm d}p \, W_{\hat{A}}^{(g)}(x,p) \hat{\Delta}^{(g)}(x,p).
\end{equation}
As desired, expectation values are calculated by the phase space integral over Wigner function and Weyl symbol,
\begin{equation}
\langle \hat{A} \rangle = {\rm tr}[\hat{\rho} \hat{A}] = \int {\rm d}x \int {\rm d}p \, W_{\hat{A}}^{(g)}(x,p) W^{(g)}(x,p).
\end{equation}
The Weyl symbols are covariant under phase space translations in the sense that the Weyl symbol of a translated operator $\hat{A}'=\hat{D}^{(g)}(x',p')\hat{A}\hat{D}^{(g) \dagger}(x',p')$ is the shifted Weyl symbol of the operator $\hat{A}$,
\begin{equation} \label{Curved_Weyl_covariance}
W_{\hat{A}'}^{(g)}(x,p) = W_{\hat{A}}^{(g)}(x-x',p-p').
\end{equation}
This directly follows from our definition (\ref{Curved_Weyl_symbol}) of the Weyl symbols and holds for arbitrary curved configuration spaces. The definition of the star product remains unchanged in curved configuration spaces,
\begin{align}
W_{\hat{A} \hat{B}}^{(g)}(x,p) =& (W_{\hat{A}}^{(g)} \star W_{\hat{B}}^{(g)})(x,p) \\
 =& \int \frac{{\rm d}x_1 {\rm d}p_1}{(\pi \hbar)^n} \int \frac{{\rm d}x_2 {\rm d}p_2}{(\pi \hbar)^n} {\rm e}^{2 {\rm i} (x_1^i p_{2 i}-x_2^i p_{1 i})/\hbar} \nonumber \\
 & \times W_{\hat{A}}^{(g)}(x+x_1, p+p_1) W_{\hat{B}}^{(g)}(x+x_2, p+p_2). \nonumber
\end{align}
Accordingly, the alternative representations (\ref{Cartesian_differential_star_product}) hold also in the general case.

The Weyl symbols of the position and momentum operators $\hat{x}^i$ and $\hat{p}_i$ take their classical phase space analogues
\begin{equation}
W_{\hat{x}^i}^{(g)}(x,p) = x^i \,\,\, \text{and} \,\,\, W_{\hat{p}_i}^{(g)}(x,p) = p_i,
\end{equation}
and the same holds for any Weyl-ordered product,
\begin{equation}
W_{\{(\hat{p}_{i})^k, (\hat{x}^{i})^{\ell}\}_{\rm W}}^{(g)}(x,p) = (p_{i})^k (x^{i})^{\ell}.
\end{equation}
The Weyl ordering is defined in Eq.~(\ref{Weyl_ordering}), and there is no summation over $i$.

The Hamiltonian (\ref{CurvedSpaceHamiltonian}) is represented by the Weyl symbol
\begin{equation} \label{Curved_Weyl_Hamiltonian}
W_{\hat{H}}^{(g)}(x,p) = \frac{1}{2 m} p_i g^{i j}(x) p_j + U(x).
\end{equation}
Here, the formal potential
\begin{equation} \label{Curved_Weyl_Potential}
U(x) = V(x) + \hbar^2 Q(x) + \frac{\hbar^2}{8 m} \frac{\partial^2}{\partial x^i \partial x^j} g^{i j}(x)
\end{equation}
includes the quantum potential (\ref{Curved_quantum_potential}) and
an additional metric correction term; the latter arises because the kinetic energy operator in Eq.~(\ref{CurvedSpaceHamiltonian}) is not Weyl ordered.

\subsection{Quantum Liouville equation}

We now establish the quantum Liouville equation in curved configuration spaces for a general Hamiltonian of the form (\ref{CurvedSpaceHamiltonian}). Like in the Cartesian case it has the general form
\begin{equation}
\partial_t W^{(g)}(x,p) = -\frac{{\rm i}}{\hbar} [W_{\hat{H}}^{(g)} \star W^{(g)} - W^{(g)} \star W_{\hat{H}}^{(g)}](x,p),
\end{equation}
and using the explicit Weyl symbol (\ref{Curved_Weyl_Hamiltonian}) we obtain ($L \equiv \sum_{k=1}^n \ell_k$)
\begin{widetext}
\begin{align} \label{Curved_Quantum_Liouville_Equation}
\partial_t W^{(g)}(x,p) =& - \left[ \frac{p_i}{m} \vec{\partial}_{x^j} \underset{L \, {\rm even}}{\sum_{\ell_1,\dots,\ell_n=0}^{\infty}} \frac{\partial^{L} g^{i j}(x)}{\partial^{\ell_1} x^1 \dots \partial^{\ell_n} x^n} (-1)^{\frac{L}{2}} \left(\frac{\hbar}{2}\right)^L \prod_{k=1}^n \frac{\vec{\partial}_{p_k}^{\ell_k}}{\ell_{k}!} \right] W^{(g)}(x,p) \nonumber \\
 & + \left[ \frac{1}{2 m} \left\{ p_i p_j - \left(\frac{\hbar}{2}\right)^2 \vec{\partial}_{x^i} \vec{\partial}_{x^j} \right\} \underset{L \, {\rm odd}}{\sum_{\ell_1,\dots,\ell_n=0}^{\infty}} \frac{\partial^{L} g^{i j}(x)}{\partial^{\ell_1} x^1 \dots \partial^{\ell_n} x^n} (-1)^{\frac{L-1}{2}} \left(\frac{\hbar}{2}\right)^{L-1} \prod_{k=1}^n \frac{\vec{\partial}_{p_k}^{\ell_k}}{\ell_{k}!} \right] W^{(g)}(x,p) \nonumber \\
 & + \left[ \underset{L \, {\rm odd}}{\sum_{\ell_1,\dots,\ell_n=0}^{\infty}} \frac{\partial^{L} U(x)}{\partial^{\ell_1} x^1 \dots \partial^{\ell_n} x^n} (-1)^{\frac{L-1}{2}} \left(\frac{\hbar}{2}\right)^{L-1} \prod_{k=1}^n \frac{\vec{\partial}_{p_k}^{\ell_k}}{\ell_{k}!} \right] W^{(g)}(x,p).
\end{align}
\end{widetext}
As in Eq.~(\ref{Cartesian_differential_star_product}), the arrows on top of the derivatives indicate that they act only on the Wigner function. In the case of a constant metric the second line vanishes. For a Euclidean space, $g^{i j}(x)=\delta^{i j}$, the Cartesian quantum Liouville equation (\ref{CartesianQuantumLiouvilleEquation}) is reproduced (in $n$ dimensions). We note that a related quantum Liouville equation has been derived in Ref.~\cite{Bondar2012a} based on a different definition of the Wigner function in curvilinear coordinates.

In the semiclassical limit, where only leading terms in $\hbar$ are retained, the quantum Liouville equation (\ref{Curved_Quantum_Liouville_Equation}) reduces to
\begin{align}
\partial_t W^{(g)}(x,p) =& \bigg[ -\frac{p_i}{m} g^{i j}(x) \partial_{x^j} + \frac{1}{2m} p_i p_j \frac{\partial g^{i j}(x)}{\partial x^k} \partial_{p_k} \nonumber \\
 & + \frac{\partial V(x)}{\partial x^k} \partial_{p_k} + \mathcal{O}(\hbar^2) \bigg] W^{(g)}(x,p).
\end{align}
It coincides with the classical Liouville equation in curved configuration spaces, as desired. This strongly confirms the viability of the presented phase space representation.

To summarize, we have verified that all kinematic and dynamic aspects of the Cartesian Wigner phase space formalism can be generalized to the situation of curved configuration spaces. In the following, we apply this phase-space formalism to problems characterized by curved configuration spaces. To ease notation, and since there is no risk of confusion, we will drop from now on the upper label $(g)$ indicating that we deal with a curved space.

\section{Unbounded two-dimensional curved spaces} \label{Section:Extended_two-dimensional_curved_spaces}

We proceed to consider the motion of a quantum particle in unbounded configuration spaces. The resulting conjugate momentum observables then exhibit continuous spectra. For the sake of clarity, we focus on a particle constrained to a two-dimensional surface embedded into three-dimensional Euclidean space. The surface is taken to be parametrized in Cartesian coordinates by
\begin{equation}
z = f(x,y).
\end{equation}
One may think of elliptic (or hyperbolic) paraboloids,
\begin{equation} \label{Constraint_paraboloid}
f(x,y) = c \left[ \left(\frac{x}{a}\right)^2 \pm \left(\frac{y}{b}\right)^2 \right],
\end{equation}
or a Gaussian-shaped bump in the vicinity of the origin,
\begin{equation}
f(x,y) = c \, {\rm e}^{-(x/a)^2-(y/b)^2}.
\end{equation}
For the time being, we keep the discussion general, merely requiring that $f(x,y)$ is sufficiently smooth.

If the particle is not subject to an additional potential the resulting classical Hamiltonian is determined by the kinetic energy $H = g^{i j}(x,y) p_i p_j/2 m$, $(i,j \in \{x,y\})$, where the metric coefficients are given by ($f_x \equiv \partial_x f$)
\begin{subequations}
\begin{align}
g^{x x}(x,y) &= 1-\frac{f_x^2}{1+f_x^2+f_y^2} \\
g^{y y}(x,y) &= 1-\frac{f_y^2}{1+f_x^2+f_y^2} \\
g^{x y}(x,y) &= g^{y x}(x,y) = -\frac{f_x f_y}{1+f_x^2+f_y^2}.
\end{align}
\end{subequations}
Based on these one gets the metric determinant
\begin{equation}
g(x,y) = 1+f_x^2(x,y)+f_y^2(x,y).
\end{equation}
Using Eq.~(\ref{CurvedWignerFunction}), we can now immediately write down the Wigner function in terms of the position representation of the statistical operator,
\begin{widetext}
\begin{align} \label{Two-dimensional_curved_Wigner}
W(x,y,p_x,p_y) = \frac{1}{(2 \pi \hbar)^2} \int_{-\infty}^{\infty} {\rm d}x' \int_{-\infty}^{\infty} {\rm d}y' & \frac{\left[1+f_x^2(x_+,y_+)+f_y^2(x_+,y_+)\right]^{1/4}}{\left[1+f_x^2(x_-,y_-)+f_y^2(x_-,y_-)\right]^{-1/4}} {\rm e}^{{\rm i} (p_x x'+p_y y')/\hbar} \langle x_-,y_-|\hat{\rho}|x_+,y_+\rangle.
\end{align}
Here we introduced the short hand notation $x_{\pm}=x \pm x'/2$, and likewise for $y_{\pm}$. Weyl symbols are determined analogously. 

The corresponding quantum Liouville equation can be obtained from Eq.~(\ref{Curved_Quantum_Liouville_Equation}). In case of the parabolic constraint (\ref{Constraint_paraboloid}) and confining to the region $x/a \ll 1$, $y/b \ll 1$, one obtains
\begin{align} \label{Parabolic_Quantum_Liouville_Equation}
\bigg( \partial_t + \frac{p_x}{m} & \vec{\partial}_x + \frac{p_y}{m} \vec{\partial}_y - \frac{\partial U}{\partial x} \vec{\partial}_{p_x} - \frac{\partial U}{\partial y} \vec{\partial}_{p_y} \bigg) W(x,y,p_x,p_y) \nonumber \\
= & \Bigg[ \frac{4 c^2}{a^4} \frac{p_x}{m} \vec{\partial}_x \left\{ x^2 - \frac{\hbar^2}{4} \vec{\partial}_{p_x}^2 \right\} + \frac{4 c^2}{b^4} \frac{p_y}{m} \vec{\partial}_y \left\{ y^2 - \frac{\hbar^2}{4} \vec{\partial}_{p_y}^2 \right\} \pm \frac{4 c^2}{a^2 b^2} \left( \frac{p_x}{m} \vec{\partial}_y + \frac{p_y}{m} \vec{\partial}_x \right) \left\{ x y - \frac{\hbar^2}{4} \vec{\partial}_{p_x} \vec{\partial}_{p_y} \right\} \nonumber \\
 & - \frac{4 c^2}{a^4} \left\{ p_x^2 - \frac{\hbar^2}{4} \vec{\partial}_{x}^2 \right\} \frac{x}{m} \vec{\partial}_{p_x} - \frac{4 c^2}{b^4} \left\{ p_y^2 - \frac{\hbar^2}{4} \vec{\partial}_{y}^2 \right\} \frac{y}{m} \vec{\partial}_{p_y} \mp \frac{4 c^2}{a^2 b^2} \left\{ p_x p_y - \frac{\hbar^2}{4} \vec{\partial}_{x} \vec{\partial}_{y} \right\} \left( \frac{y}{m} \vec{\partial}_{p_x} + \frac{x}{m} \vec{\partial}_{p_y} \right) \nonumber \\
 & + \underset{\ell_x+\ell_y \equiv L \, {\rm odd}}{\sum_{\ell_x,\ell_y=1}^{\infty}} \frac{\partial^L U(x,y)}{\partial x^{\ell_x} \partial y^{\ell_y}} (-1)^{\frac{L-1}{2}} \left(\frac{\hbar}{2}\right)^{L-1} \frac{\vec{\partial}_{p_x}^{\ell_x} \vec{\partial}_{p_y}^{\ell_y}}{\ell_x! \ell_y!} \; \Bigg] W(x,y,p_x,p_y).
\end{align}
\end{widetext}
The effective potential $U(x,y)$ is defined as in Eq.~(\ref{Curved_Weyl_Potential}). The explicit quantum potential $Q(x,y)$ is unwieldy; for the elliptic case and $a=b$ it reads
\begin{equation}
Q(x,y) = \frac{1}{m} \frac{a^4 [2 a^4 - 3 (x^2+y^2)]}{2 [a^4 + 2 (x^2+y^2)]^3}.
\end{equation}
The left-hand side of Eq.~(\ref{Parabolic_Quantum_Liouville_Equation}) describes the classical evolution of a point particle in Euclidean space subject to the potential $U$, while the second and the third line capture curvature-induced effects. The last line describes the familiar quantum corrections due to anharmonic potentials. If one approximates the metric  by a Taylor series up to second order (e.g.~in the vicinity of an extremal point) the resulting quantum Liouville equation has the same structure (\ref{Parabolic_Quantum_Liouville_Equation}).

\begin{figure}
\includegraphics[width=\linewidth]{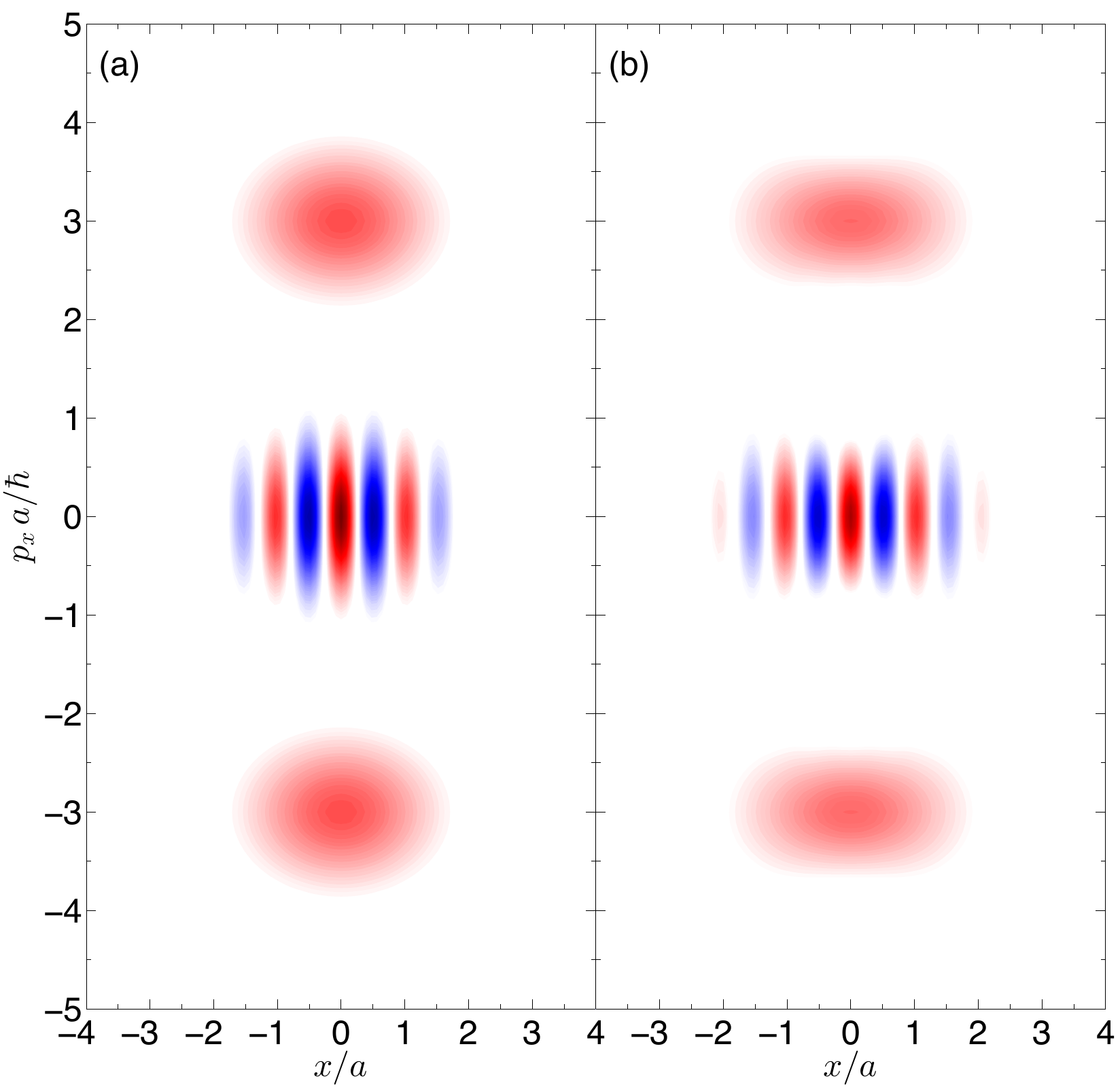}
\caption{\label{Fig1} (Color online) Wigner function of a superposition state in flat (left) and curved (right) two-dimensional configuration space; a slice of the four-dimensional Wigner function (\ref{Two-dimensional_curved_Wigner}) is shown for $y=p_y=0$. The curved space is taken to be an elliptic paraboloid, Eq.~(\ref{Constraint_paraboloid}), with $a/c=b/c=1$. The quantum state is a superposition of two Gaussian wave packets with mean momenta $(\overline{p}_x, \overline{p}_y) = (\pm 3 \hbar/a, 0) $, both centered at the origin with widths $\sigma_{x,y}=a$. Red [light gray] (blue [dark gray]) color indicates positive (negative) values of the density plot, normalized to the maximal value.}
\end{figure}

To illustrate the effect of the curvature, we display in Fig.~\ref{Fig1} the Wigner function of a superposition of two Gaussian wave packets with different velocities, both centered around the origin. Comparing the Wigner functions on a flat plane (left) and on a elliptic paraboloid (right), one observes that the characteristic structure of the superposition state is preserved in the curved case. The main effect of the curvature is to distort the envelopes of the individual components and the resulting interference.

\section{Quantum particle on a sphere} \label{Section:Free_particle_on_a_sphere}

In the following, we demonstrate how the phase space representation applies to compact configuration spaces. To this end, we consider a particle constrained to the surface of a sphere, one of the paradigms for quantum mechanics in curved space. The underlying compact symmetry group ${\rm SO}(3)$ exhibits finite-dimensional representations, e.g.~the spherical harmonics for fixed total angular momentum quantum number $\ell$. The motion on a sphere has been widely investigated in the literature, e.g.~in the context of establishing a phase-space formalism for the spin degree of freedom (based on the SU(2) algebra) \cite{Varilly1989a, Brif1999a, Alonso2003a, Klimov2008a, Carinena2012a}.

\subsection{Phase space coordinates on the sphere}

The appropriate generalized coordinates to describe the dynamics on a sphere are the azimuthal angle $\varphi \in [0,2 \pi)$ and the polar angle $\vartheta \in [0,\pi]$ in spherical coordinates, where the radial coordinate $r$ is constrained to fixed radius $r=R$. The classical Hamilton function for a particle of mass $M$ moving freely on this surface is then given by
\begin{equation} \label{Classical_Hamiltonian_Sphere}
H = \frac{1}{2 M R^2} \left(p_{\vartheta}^2 + \frac{p_{\varphi}^2}{\sin^2 \vartheta} \right).
\end{equation}
Here $p_{\varphi} = \vec{e}_z \cdot \vec{L}$ and $p_{\vartheta} = - \vec{e}_{\varphi} \cdot \vec{L}$ describe the components of the angular momentum $\vec{L}$ with respect to the unit vectors $\vec{e}_z$ and $\vec{e}_{\varphi}$. Comparing Eq.~(\ref{Classical_Hamiltonian_Sphere}) with Eq.~(\ref{Classical_Hamiltonian_curved}), i.e.~with the general form $H=g^{i j}(\vartheta, \varphi) p_i p_j / 2 M R^2$, ($i,j \in \{\vartheta, \varphi\}$), we can identify the metric components
\begin{subequations}
\begin{align} \label{Metric_sphere}
g^{\vartheta \vartheta}(\vartheta, \varphi) &= 1 \\
g^{\varphi \varphi}(\vartheta, \varphi) &= \frac{1}{\sin^2 \vartheta} \\
g^{\vartheta \varphi}(\vartheta, \varphi) &= g^{\varphi \vartheta}(\vartheta, \varphi) = 0.
\end{align}
\end{subequations}
From this one infers the covariant components $g_{\vartheta \vartheta}(\vartheta, \varphi)=1$, $g_{\varphi \varphi}(\vartheta, \varphi)=\sin^2 \vartheta$ and $g_{\vartheta \varphi}(\vartheta, \varphi) = g_{\varphi \vartheta}(\vartheta, \varphi) = 0$, yielding the metric determinant
\begin{equation} \label{Metric_determinant_sphere}
g(\vartheta, \varphi) = \left| \begin{array}{cc} 1 & 0 \\ 0 & \sin^2 \vartheta \end{array} \right| = \sin^2 \vartheta.
\end{equation}
According to Eq.~(\ref{Coordinate_basis_curved}), the identity operator in the coordinate basis $|\vartheta, \varphi \rangle$ is therefore given by
\begin{equation}
\mathbb{1} = \int_0^{\pi} {\rm d}\vartheta \int_0^{2 \pi} {\rm d}\varphi \, \sin \vartheta \, |\vartheta, \varphi \rangle \langle \vartheta, \varphi |.
\end{equation}
Moreover, the conjugate momentum operators follow from Eq.~(\ref{CurvedSpaceMomentumQuantizationRule}):
\begin{equation} \label{Canonically_conjugate_momentum_operators_sphere}
\hat{p}_{\varphi} = \frac{\hbar}{\rm i} \frac{\partial}{\partial \varphi}, \qquad \hat{p}_{\vartheta} = \frac{\hbar}{\rm i} \left( \frac{\partial}{\partial \vartheta} + \frac{1}{2} \cot \vartheta \right).
\end{equation}
The finite range of the angular coordinates introduces a novel feature, the occurrence of discrete momentum eigenvalues. The eigenfunctions are given by (cf.~Eq.~(\ref{Momentum_eigenstates_curved}))
\begin{equation}
\langle \vartheta, \varphi | m_{\vartheta}, m_{\varphi} \rangle = \frac{\mathrm{e}^{2 {\rm i} m_{\vartheta} \vartheta}}{\sqrt{\pi \sin \vartheta}} \frac{\mathrm{e}^{{\rm i} m_{\varphi} \varphi}}{\sqrt{2 \pi}},
\end{equation}
with $m_{\varphi}, m_{\vartheta} \in \mathbb{Z}$ labeling the associated eigenvalues,
\begin{subequations}
\begin{align}
\hat{p}_\vartheta |m_{\vartheta}, m_{\varphi}\rangle &= 2 \hbar m_{\vartheta} |m_{\vartheta}, m_{\varphi}\rangle, \\
\hat{p}_\varphi |m_{\vartheta}, m_{\varphi}\rangle &= \hbar m_{\varphi} |m_{\vartheta}, m_{\varphi}\rangle.
\end{align}
\end{subequations}
The eigenvectors form a discrete orthonormal basis of the Hilbert space,
\begin{subequations}
\begin{align}
& \mathbb{1} = \sum_{m_{\vartheta} \in \mathbb{Z}} \sum_{m_{\varphi} \in \mathbb{Z}} |m_{\vartheta}, m_{\varphi}\rangle \langle m_{\vartheta}, m_{\varphi}|, \\
& \langle m_{\vartheta}, m_{\varphi} | m_{\vartheta}', m_{\varphi}' \rangle = \delta_{m_{\vartheta},m_{\vartheta}'} \delta_{m_{\varphi},m_{\varphi}'}.
\end{align}
\end{subequations}

With this we have identified the phase space. Unlike in the unbounded configuration space the discrete momentum spectrum implies a discrete momentum phase space coordinate. This is a generic feature of compact coordinates and also arises, e.g., in the case of a single angle variable (motion on a circle) \cite{Rigas2011a} or the orientation state of a rigid body \cite{Fischer2013a}. We stress that this discreteness does not arise due to the phase space formalism, but is a necessary physical consequence, as also reflected in the discrete measurement outcomes of the corresponding momentum observables. We see below that the classical continuous momentum space is regained in the semiclassical limit.

We conclude with a number of further remarks.

\subsubsection*{Discussion}

First, the commutators $[\hat{\varphi},\hat{p}_{\varphi}]$ and $[\hat{\vartheta},\hat{p}_{\vartheta}]$ are operator valued, which is a generic feature of position and momentum operator pairs in compact spaces that is not relevant in the following. In particular, it does not affect the conjugate relationship between the angle coordinates and the momentum operators (\ref{Canonically_conjugate_momentum_operators_sphere}), which is founded on the classical Poisson brackets. The necessity of operator-valued commutators can be seen from the uncertainty relation, as the bounded coordinate uncertainty can give rise to vanishing uncertainty products, e.g.~in the case of momentum eigenstates. We remind the reader that the displacement operators (\ref{CartesianDisplacementOperators}) and (\ref{Curved_Displacement_Operators}) are not defined symmetrically, because one cannot apply the Baker-Campbell-Hausdorff formula if the commutator is operator valued.

As a second remark, one might suppose that the spherical harmonics offer an alternative, viable momentum basis to complement the angle coordinates:
\begin{subequations}
\begin{align}
\mathbb{1} &= \sum_{\ell=0}^{\infty} \sum_{m=-\ell}^{\ell} |\ell m\rangle \langle \ell m|, \\
 &\langle \vartheta,\varphi|\ell m\rangle = Y_{\ell,m}(\vartheta,\varphi).
\end{align}
\end{subequations}
The momentum operators would then be the total angular momentum $\hat{\vec{L}}^2$ and its projection on the $z$-axis $\hat{L}_z$. This choice seems in particular appealing, since the spherical harmonics, realizing a representation of the rotation group SO(3), respect the continuous symmetry of the configuration space. Moreover, a free particle constrained to the surface of a sphere is described by the Hamiltonian
\begin{equation} \label{Spherical_harmonics_Hamiltonian_sphere}
\hat{H} = \frac{\hat{\vec{L}}^2}{2 M R^2} = \sum_{\ell=0}^{\infty} \sum_{m=-\ell}^{\ell} \frac{\hbar^2 \ell (\ell+1)}{2 M R^2} |\ell m\rangle \langle \ell m|.
\end{equation}
The time evolution of an arbitrary quantum state can thus easily be given in terms of the spherical harmonics.

However, for this choice it is not possible to relate each coordinate to a corresponding momentum, impeding a phase-space formalism that satisfies the marginal property. Next, even though $\hat{\vec{L}}^2$ and $\hat{L}_z$ commute, they do not constitute independent variables, as can be seen from the range dependence of the quantum number $m$ on $\ell$. Moreover, since they are not the momenta conjugate to the coordinates, they will not yield the desired transition to the classical Liouville equation in the semiclassical limit. All this illustrates that the issue how to establish a viable phase-space formalism is independent of possible symmetries of the configuration space.

\subsection{Wigner-Weyl representation on the sphere}

Even though the spectra of the conjugate momenta are discrete, the derivation of the Wigner-Weyl representation follows by and large the same line of argument as in Section \ref{Section:Curved_space_Wigner_function}. All properties and relations of the Wigner-Weyl formalism carry over, except that the continuous momentum argument is replaced by a discrete variable and the momentum integrals by summations.

We introduce the displacement operators $\hat{D}(\vartheta,\varphi,m_{\vartheta},m_{\varphi}) = \hat{D}_{\vartheta}(\vartheta,m_{\vartheta}) \hat{D}_{\varphi}(\varphi,m_{\varphi})$ for $\vartheta \in [0,\pi]$, $\varphi \in [0, 2 \pi)$, and $m_{\vartheta},m_{\varphi} \in \mathbb{Z}$ in terms of the commuting factors
\begin{subequations} \label{Displacement_operators_sphere}
\begin{align}
\hat{D}_{\vartheta}(\vartheta,m_{\vartheta}) &= {\rm e}^{2 {\rm i} m_{\vartheta} \hat{\vartheta}} {\rm e}^{-{\rm i} \vartheta \hat{p}_{\vartheta}/\hbar}, \\
\hat{D}_{\varphi}(\varphi,m_{\varphi}) &= {\rm e}^{{\rm i} m_{\varphi} \hat{\varphi}} {\rm e}^{-{\rm i} \varphi \hat{p}_{\varphi}/\hbar}.
\end{align}
\end{subequations}
Note that the translation operators keep the phase-space variables within their range of definition:
\begin{subequations}
\begin{align}
{\rm e}^{-{\rm i} \vartheta' \hat{p}_{\vartheta}/\hbar} |\vartheta,\varphi \rangle = &\sqrt{\frac{|\sin(\vartheta+\vartheta')|}{\sin \vartheta}} |[\vartheta+\vartheta'] \, {\rm mod} \, \pi,\varphi \rangle \\
{\rm e}^{{2 \rm i} m_{\vartheta}' \hat{\vartheta}} |m_{\vartheta},m_{\varphi} \rangle &= |m_{\vartheta}+m_{\vartheta}',m_{\varphi} \rangle \\
{\rm e}^{-{\rm i} \varphi' \hat{p}_{\varphi}/\hbar} |\vartheta,\varphi \rangle &= |\vartheta, [\varphi+\varphi'] \, {\rm mod} \, 2 \pi \rangle \\
{\rm e}^{{\rm i} m_{\varphi}' \hat{\varphi}} |m_{\vartheta},m_{\varphi} \rangle &= |m_{\vartheta},m_{\varphi}+m_{\varphi}' \rangle.
\end{align}
\end{subequations}
The undisplaced operator $\hat{\Delta}_0=\hat{\Delta}_{\vartheta}(0,0) \hat{\Delta}_{\varphi}(0,0)$ is defined by
\begin{subequations}
\begin{align}
\hat{\Delta}_{\vartheta}(0,0) =& \frac{1}{\pi} \sum_{m_{\vartheta}' \in \mathbb{Z}} \int_{-\pi/2}^{\pi/2}{\rm d}\vartheta' \hat{D}_{\vartheta}(\vartheta',m_{\vartheta}') {\rm e}^{-{\rm i} \vartheta' m_{\vartheta}'}, \\
\hat{\Delta}_{\varphi}(0,0) =& \frac{1}{2 \pi} \sum_{m_{\varphi}' \in \mathbb{Z}} \int_{-\pi}^{\pi}{\rm d}\varphi' \hat{D}_{\varphi}(\varphi',m_{\varphi}') {\rm e}^{-{\rm i} \varphi' m_{\varphi}'/2}.
\end{align}
\end{subequations}
The main difference with the general undisplaced kernel (\ref{Curved_undisplaced_Stratonovich_Weyl_operator_kernel}) is that the momentum integrals are replaced by sums. Moreover, the angle integrations are defined symmetrically with respect to the origin. This is required in order to guarantee the hermiticity of $\hat{\Delta}_0$. It is not in conflict with the definition ranges of the angle variables, since the integration variables correspond to changes of the angles.

\begin{figure}
\includegraphics[width=\linewidth]{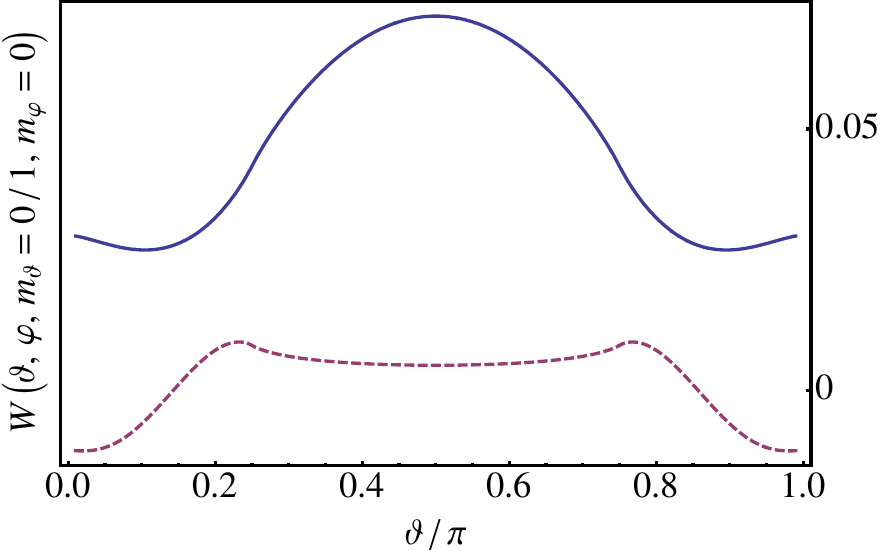}
\caption{\label{Fig2} Wigner function $W(\vartheta, \varphi, m_{\vartheta}, m_{\varphi})$ of the isotropic spherical harmonic $|\ell=0,m=0\rangle$. Shown is the variation with the polar angle $\vartheta$ for the conjugate momentum choices $m_{\vartheta}=0$ (solid line) and $m_{\vartheta}=1$ (dashed line); in both cases we specify $m_{\varphi}=0$. The Wigner function is independent of the azimuthal angle $\varphi$.}
\end{figure}

The general Stratonovich-Weyl operator kernel $\hat{\Delta}(\vartheta,\varphi,m_{\vartheta},m_{\varphi})$ is defined analogously to (\ref{Curved_Stratonovich_Weyl_quantizer_definition}) and reads
\begin{align}
\hat{\Delta}(\vartheta,\varphi,m_{\vartheta},m_{\varphi}) =& \int_{-\pi/2}^{\pi/2}{\rm d}\vartheta' \int_{-\pi}^{\pi}{\rm d}\varphi' \sqrt{\sin \vartheta_+ \sin \vartheta_-} \nonumber \\
 & \times {\rm e}^{2 {\rm i} m_{\vartheta} \vartheta'} {\rm e}^{{\rm i} m_{\varphi} \varphi'} |\vartheta_+,\varphi_+\rangle \langle \vartheta_-,\varphi_-|.
\end{align}
Here we introduced the abbreviations $\vartheta_{\pm}=[\vartheta \pm \vartheta'/2] {\rm mod} \, \pi$ and $\varphi_{\pm}=[\varphi \pm \varphi'/2] {\rm mod} \, 2 \pi$. The operator kernel satisfies the completeness relation
\begin{align}
{\rm tr}&[\hat{\Delta}(\vartheta,\varphi,m_{\vartheta},m_{\varphi}) \hat{\Delta}(\vartheta',\varphi',m_{\vartheta}',m_{\varphi}')] \\
=& 2 \pi^2 \delta_{m_{\vartheta},m_{\vartheta}'} \delta([\vartheta-\vartheta'] {\rm mod} \, \pi) \delta_{m_{\varphi},m_{\varphi}'} \delta([\varphi-\varphi'] {\rm mod} \, 2 \pi). \nonumber
\end{align}
The Wigner function for a point particle constrained to the surface of a sphere is now given by
\begin{align} \label{Wigner_function_sphere}
&W(\vartheta,\varphi,m_{\vartheta},m_{\varphi}) = \frac{1}{2 \pi^2} {\rm tr}[\hat{\rho} \hat{\Delta}(\vartheta,\varphi,m_{\vartheta},m_{\varphi})] \nonumber \\
&= \frac{1}{2 \pi^2} \int_{-\pi/2}^{\pi/2}{\rm d}\vartheta' \int_{-\pi}^{\pi}{\rm d}\varphi' \sqrt{\sin \vartheta_+ \sin \vartheta_-} \nonumber \\
 &\phantom{=} \times {\rm e}^{2 {\rm i} m_{\vartheta} \vartheta'} {\rm e}^{{\rm i} m_{\varphi} \varphi'} \langle \vartheta_-,\varphi_-|\hat{\rho}|\vartheta_+,\varphi_+\rangle;
\end{align}
it corresponds to Eq. (\ref{CurvedWignerFunction}) with modified integration limits. The Weyl symbol of an operator $\hat{A}$ is defined accordingly:
\begin{align}
&W_{\hat{A}}(\vartheta,\varphi,m_{\vartheta},m_{\varphi}) = {\rm tr}[\hat{A} \hat{\Delta}(\vartheta,\varphi,m_{\vartheta},m_{\varphi})] \nonumber \\
&= \int_{-\pi/2}^{\pi/2}{\rm d}\vartheta' \int_{-\pi}^{\pi}{\rm d}\varphi' \sqrt{\sin \vartheta_+ \sin \vartheta_-} \nonumber \\
 & \phantom{=} \times {\rm e}^{2 {\rm i} m_{\vartheta} \vartheta'} {\rm e}^{{\rm i} m_{\varphi} \varphi'} \langle \vartheta_-,\varphi_-|\hat{A}|\vartheta_+,\varphi_+\rangle.
\end{align}
The inverse relation (\ref{Curved_Weyl_inverse}) holds with the integral over $p$ replaced by summations over $m_{\vartheta}$ and $m_{\varphi}$.  For the momenta $\hat{p}_{\vartheta}$ and $\hat{p}_{\varphi}$ one now obtains the desired Weyl symbols $W_{\hat{p}_{\vartheta}}(\vartheta,\varphi,m_{\vartheta},m_{\varphi}) = 2 \hbar m_{\vartheta}$ and $W_{\hat{p}_{\varphi}}(\vartheta,\varphi,m_{\vartheta},m_{\varphi}) = \hbar m_{\varphi}$. The marginals evaluate as expected,
\begin{subequations}
\begin{align}
\int_0^{\pi} {\rm d}\vartheta \int_0^{2 \pi} {\rm d}\varphi \, W(\vartheta,\varphi,m_{\vartheta},m_{\varphi}) &= \langle m_{\vartheta},m_{\varphi}|\hat{\rho}|m_{\vartheta},m_{\varphi} \rangle, \\
\sum_{m_{\vartheta},m_{\varphi} \in \mathbb{Z}} W(\vartheta,\varphi,m_{\vartheta},m_{\varphi}) &= \sqrt{\sin \vartheta} \langle \vartheta,\varphi|\hat{\rho}|\vartheta,\varphi  \rangle.
\end{align}
\end{subequations}

As an example, we consider the Wigner function for the isotropic angular momentum state $|\ell =0,m=0\rangle$ with $\langle \vartheta, \varphi|\ell =0, m=0\rangle = 1/\sqrt{4 \pi}$. It simplifies to an expression
\begin{align} \label{Wigner_function_s-wave}
W(\vartheta, \varphi, m_{\vartheta}, m_{\varphi}) =& \frac{1}{4 \pi^2} \delta_{m_{\varphi},0} \int_{-\pi/2}^{\pi/2} {\rm d}\vartheta' \sqrt{|\sin(\vartheta+\vartheta'/2)|} \nonumber \\
& \times \sqrt{|\sin(\vartheta-\vartheta'/2)|} \cos(2 m_{\vartheta} \vartheta'),
\end{align}
which is independent of the azimuthal angle variable $\varphi$, but depends on the polar angle variable $\vartheta$ and the conjugate momenta $m_{\vartheta}$ and $m_{\varphi}$. This is illustrated in Fig.~\ref{Fig2} for the cuts $W(\vartheta, \varphi, m_{\vartheta}=0, m_{\varphi}=0)$ and $W(\vartheta, \varphi, m_{\vartheta}=1, m_{\varphi}=0)$. The fact that an isotropic state exhibits angular dependence is a consequence of the use of spherical coordinates, which are defined with respect to a chosen direction.

We have thus established that the Wigner function for a particle on a sphere behaves like a quasi-probability distribution. We remark that the Wigner function (\ref{Wigner_function_sphere}) is not invariant under rotations, in the sense that the Weyl symbol $W_{R_{\rm H} \hat{\rho} R_{\rm H}^{\dagger}}(x)$ of a rotated state $\hat{R}_{\rm H} \hat{\rho} \hat{R}_{\rm H}^{\dagger}$ is not the rotated Weyl symbol $W_{\hat{\rho}}(R_{\rm C} x)$ of the unrotated state (H and C indicate that the rotation acts in Hilbert or in phase space, respectively). This is evident by considering the Wigner function of an isotropic state such as Eq.~(\ref{Wigner_function_s-wave}). While the state is invariant under rotations, this does not hold for the phase space representation due to its angular dependence. This consequence of the choice of spherical coordinates arises also in the classical case. As in the general case (\ref{Curved_Weyl_covariance}), the Weyl symbols are covariant under the displacements (\ref{Displacement_operators_sphere}), instead.

The momentum representation of the Wigner function is more intricate due to the discreteness of the momenta, and reads
\begin{widetext}
\begin{align}
W(\vartheta,\varphi,m_{\vartheta},m_{\varphi}) =& \frac{1}{2 \pi^2} \sum_{m_{\vartheta}',m_{\vartheta}'',m_{\varphi}',m_{\varphi}'' \in \mathbb{Z}} {\rm sinc} \left[ \left( m_{\vartheta} - \frac{m_{\vartheta}'+m_{\vartheta}''}{2} \right) \pi \right] {\rm sinc} \left[ \left( m_{\varphi} - \frac{m_{\varphi}'+m_{\varphi}''}{2} \right) \pi \right] \nonumber \\
 & \times {\rm e}^{2 {\rm i} (m_{\vartheta}'-m_{\vartheta}'') \vartheta} {\rm e}^{{\rm i} (m_{\varphi}'-m_{\varphi}'') \varphi} \langle m_{\vartheta}',m_{\varphi}'|\hat{\rho}|m_{\vartheta}'',m_{\varphi}'' \rangle,
\end{align}
\end{widetext}
with ${\rm sinc}(x)=\sin(x)/x$. We remark that one could obtain a diagonal momentum representation with only two sums by choosing an undisplaced operator kernel $\hat{\Delta}_0$ which, however, does not preserve parity. For instance, one can take the kernel $\hat{\Delta}_{\vartheta}(0,0) = \hat{P}_{\vartheta}+{\rm e}^{\pm 2 {\rm i} \hat{\vartheta}} \hat{P}_{\vartheta}$, with $\hat{P}_{\vartheta}=\sum_{m_{\vartheta} \in \mathbb{Z}} |m_{\vartheta}\rangle \langle -m_{\vartheta}|$, and similarly for $\hat{\Delta}_{\varphi}(0,0)$. This possibility is discussed in detail in Ref.~\cite{Rigas2011a}.

\subsection{Quantum Liouville equation on the sphere}

The Hamiltonian (\ref{Spherical_harmonics_Hamiltonian_sphere}) rewritten in terms of the canonical phase space coordinates reads \cite{DeWitt1952a}
\begin{equation} \label{Quantum_Hamiltonian_Sphere}
\hat{H} = \frac{1}{2 M R^2} \left(\hat{p}_{\vartheta}^2 + \frac{\hat{p}_{\varphi}^2}{\sin^2 \hat{\vartheta}} \right) - \frac{\hbar^2}{8 M R^2} \left( 1+\frac{1}{\sin^2 \hat{\vartheta}} \right).
\end{equation}
By comparison with Eq.~(\ref{CurvedSpaceHamiltonian}) and the classical Hamiltonian (\ref{Classical_Hamiltonian_Sphere}), we can infer the quantum potential
\begin{equation}
Q(\vartheta) = - \frac{1}{8 M R^2} \left( 1+\frac{1}{\sin^2 \vartheta} \right),
\end{equation}
which agrees with the general theory (\ref{Curved_quantum_potential}).
The effect of the quantum potential is thus to repel the particle from the range boundaries $0$ and $\pi$ of the angle $\vartheta$.

Because of the discrete momenta, we cannot use Eq.~(\ref{Curved_Quantum_Liouville_Equation}) for obtaining the quantum Liouville equation. Instead, one must reevaluate the von Neumann equation with Eq.~(\ref{Quantum_Hamiltonian_Sphere}) in phase space for the discrete momentum variables. A lengthy calculation yields
\begin{widetext}
\begin{align} \label{Quantum_Liouville_equation_sphere}
\bigg( \partial_t &+ \frac{2 \hbar m_{\vartheta}}{M R^2} \partial_{\vartheta} \bigg) W(\vartheta,\varphi,m_{\vartheta},m_{\varphi}) \nonumber \\
=& - \frac{\hbar m_{\varphi}}{M R^2} \partial_{\varphi} \sum_{n=0}^{\infty} \frac{\partial^{2 n} g^{\varphi \varphi}(\vartheta,\varphi)}{\partial \vartheta^{2 n}} (-1)^{n} \frac{(\hbar/2)^{2 n}}{(2 n)!} \frac{1}{(2 \hbar)^{2 n}} \sum_{m_{\vartheta}' \in \mathbb{Z}} \delta_{m_{\vartheta}-m_{\vartheta}'}^{(2 n)} W(\vartheta,\varphi,m_{\vartheta}',m_{\varphi}) \nonumber \\
 & + \frac{1}{2 M R^2} \left\{ (\hbar m_{\varphi})^2 - \left( \frac{\hbar}{2} \right)^2 \partial_{\varphi}^2 \right\} \sum_{n=0}^{\infty} \frac{\partial^{2 n+1} g^{\varphi \varphi}(\vartheta,\varphi)}{\partial \vartheta^{2 n+1}} (-1)^{n} \frac{(\hbar/2)^{2 n}}{(2 n+1)!} \frac{1}{(2 \hbar)^{2 n+1}} \sum_{m_{\vartheta}' \in \mathbb{Z}} \delta_{m_{\vartheta}-m_{\vartheta}'}^{(2 n+1)} W(\vartheta,\varphi,m_{\vartheta}',m_{\varphi}) \nonumber \\
 & + \sum_{n=0}^{\infty} \frac{\partial^{2 n+1} Q(\vartheta)}{\partial \vartheta^{2 n+1}} (-1)^{n} \frac{(\hbar/2)^{2 n}}{(2 n+1)!} \frac{1}{(2 \hbar)^{2 n+1}} \sum_{m_{\vartheta}' \in \mathbb{Z}} \delta_{m_{\vartheta}-m_{\vartheta}'}^{(2 n+1)} W(\vartheta,\varphi,m_{\vartheta}',m_{\varphi}) \nonumber \\
 & -\frac{2\hbar(-1)^{m_\vartheta}}{MR^2\pi} \partial_\vartheta A(\vartheta,\varphi;m_\varphi)
 -\frac{\hbar (-1)^{m_\varphi}}{M R^2 \pi} \sum_{n=0}^{\infty} \frac{\partial^{2 n} g^{\varphi \varphi}(\vartheta, \varphi)}{\partial \vartheta^{2 n}} (-1)^n \frac{(\hbar/2)^{2n}}{(2 n)!} \frac{1}{(2 \hbar)^{2 n}} \sum_{m_{\vartheta}' \in \mathbb{Z}} \delta_{m_{\vartheta}-m_{\vartheta}'}^{(2 n)} \partial_\varphi B(\vartheta,\varphi;m_\vartheta') \nonumber \\
 & +\frac{\hbar (-1)^{m_\varphi}}{2 M R^2 \pi} \sum_{n=0}^{\infty} \frac{\partial^{2 n+1} g^{\varphi \varphi}(\vartheta, \varphi)}{\partial \vartheta^{2 n+1}} (-1)^n \frac{(\hbar/2)^{2n}}{(2 n+1)!} \frac{1}{(2 \hbar)^{2 n+1}} \sum_{m_{\vartheta}' \in \mathbb{Z}} \delta_{m_{\vartheta}-m_{\vartheta}'}^{(2 n+1)} \big[\hbar m_{\varphi} B(\vartheta,\varphi;m_\vartheta)+ C(\vartheta,\varphi;m_\vartheta')\big],
\end{align}
\end{widetext}
where we introduced $\delta_m^{(N)} = \partial_m^{N} {\rm sinc}(m \pi)$.\footnote{The last two lines of Eq.~(\ref{Quantum_Liouville_equation_sphere}) are missing in the originally published article Phys.~Rev.~A~{88}, 062117 (2013). They are described in the erratum Phys.~Rev.~A~{106}, 069904(E) (2022).} The metric component $g^{\varphi \varphi}(\vartheta,\varphi) = 1/\sin^2 \vartheta$ was given in Eq.~(\ref{Metric_sphere}). Note that Eq.~(\ref{Quantum_Liouville_equation_sphere}) does not contain a potential $V(\vartheta, \varphi)$, which, however, is easy to include by replacing $Q$ by $V+Q$.

Equation (\ref{Quantum_Liouville_equation_sphere}) involves the limits
\begin{subequations}
\begin{align}\label{limits}
A(\vartheta,\varphi;m_\varphi)&=\lim_{\vartheta'\uparrow\frac{\pi}{2}}\sum_{m_\vartheta'\in\mathbb{Z}}\sin(2m_\vartheta'\vartheta')
W(\vartheta,\varphi,m_\vartheta',m_\varphi),\nonumber \\
B(\vartheta,\varphi;m_\vartheta')&=\lim_{\varphi'\uparrow\pi}\sum_{m_\varphi'\in\mathbb{Z}}\sin(m_\varphi'\varphi')  W(\vartheta,\varphi,m_\vartheta',m_\varphi'),\nonumber \\
C(\vartheta,\varphi;m_\vartheta')&=\lim_{\varphi'\uparrow\pi}\sum_{m_\varphi'\in\mathbb{Z}} \hbar m_{\varphi}' \sin(m_\varphi'\varphi') W(\vartheta,\varphi,m_\vartheta',m_\varphi').\nonumber
\end{align}
\end{subequations}
These expressions do not vanish in general. This is because the convergence of the series may not be absolute so that one is not necessarily allowed to interchange the limits and the summations. Indeed, for quantum states displaying spatial coherences between antipodal positions they may yield a finite contribution. On the other hand, the limits vanish for sufficiently classical states, whose angular coherences $c_{\varphi,\vartheta}(\varphi',\vartheta')=\langle \varphi_-,\vartheta_-|\rho| \varphi_+,\vartheta_+\rangle$ decay sufficiently fast, so that they do not `feel' the fact that they are living on a finite configuration space: The last two lines in (\ref{Quantum_Liouville_equation_sphere}) can be omitted if $c_{\varphi,\vartheta}=0$ for $|\varphi'|>\pi-\epsilon$ and $|\vartheta'|>\pi/2-\epsilon$ with $\epsilon>0$, admitting the following semiclassical approximation.

If the Wigner function and its derivatives vary sufficiently slowly as functions of $m_{\vartheta}$, we can approximate the sums over $m_{\vartheta}$ by integrals. Rewriting $2 \hbar m_{\vartheta}=p_{\vartheta}$, we have
\begin{equation}
\frac{1}{(2 \hbar)^{N}} \sum_{m_{\vartheta}' \in \mathbb{Z}} \delta_{m_{\vartheta}-m_{\vartheta}'}^{(N)} W(m_{\vartheta}') \approx \partial_{p_{\vartheta}}^{N} W(p_{\vartheta}).
\end{equation}
Based on this replacement, one can now confirm easily that the quantum Liouville equation with discrete momenta, Eq.~(\ref{Quantum_Liouville_equation_sphere}), is consistent with the general version (\ref{Curved_Quantum_Liouville_Equation}) given in Section \ref{Section:Curved_space_Wigner_function}. In particular, we find that in the semiclassical limit ($\hbar \rightarrow 0$)
\begin{equation} \label{Classical_Liouville_equation_sphere}
\partial_t W = -\frac{1}{M R^2} \left(\frac{p_{\varphi}}{\sin^2 \vartheta} \partial_\varphi + p_{\vartheta} \partial_\vartheta + \frac{\cot \vartheta}{\sin^2 \vartheta} \, p_{\varphi}^2 \partial_{p_{\vartheta}} \right) W.
\end{equation}
As expected, this equation corresponds to the classical Liouville equation for a particle on the sphere. (Here we also replaced $\hbar m_{\varphi} = p_{\varphi}$.)

\subsection{Extension to the orientation state}

Finally, let us briefly sketch how the phase space representation of the motion on a sphere can be generalized to the orientation state of a rigid body. The latter can be characterized by the Euler angles $\alpha,\beta,\gamma$ of precession, nutation, and intrinsic rotation, which constitute a compact, curved configuration space. The orientation state is related to the point particle constrained to a sphere, since the two angles $\alpha,\beta$ locate the intersection point of the intrinsic rotation axis with the surface of the unit sphere; the third angle $\gamma$ describes the rotation about this axis.

The derivation of the orientation state Wigner function, as well as the quantum Liouville equation in the semiclassical limit, were already presented in a previous publication \cite{Fischer2013a}. Here we remark only that these results can also be derived in the spirit of a curved configuration space. To this end, we consider the Hamilton function of a general, non-symmetric top with the moments of inertia $I_1$, $I_2$ and $I_3$,
\begin{align}
H = & \frac{1}{2 I_1 \sin^2 \beta} \left( (p_{\alpha}-p_{\gamma} \cos \beta) \cos \gamma - p_{\beta} \sin \beta \sin \gamma \right)^2 \nonumber \\
 & + \frac{1}{2 I_2 \sin^2 \beta} \left( (p_{\alpha}-p_{\gamma} \sin \beta) \cos \gamma + p_{\beta} \sin \beta \cos \gamma \right)^2 \nonumber \\
 & + \frac{1}{2 I_3} p_{\gamma}^2.
\end{align}
Writing $H=p_i g^{i j} p_j/2 I$, $i,j \in \{ \alpha,\beta,\gamma \}$, with $I=\sqrt[3]{I_1 I_2 I_3}$, we can infer the covariant metric components
\begin{subequations} \label{Orientation_state_metric}
\begin{align}
g_{\alpha \alpha} &= \tilde{I}_1 \sin^2 \beta \cos^2 \gamma + \tilde{I}_2 \sin^2 \beta \sin^2 \gamma + \tilde{I}_3 \cos^2 \beta \\
g_{\beta \beta} &= \tilde{I}_1 \sin^2 \gamma + \tilde{I}_2 \cos^2 \gamma \\
g_{\gamma \gamma} &= \tilde{I}_3 \\
g_{\alpha \beta} &= g_{\beta \alpha} = (\tilde{I}_2-\tilde{I}_1) \sin \beta \sin \gamma \cos \gamma \\
g_{\alpha \gamma} &= g_{\gamma \alpha} = \tilde{I}_3 \cos \beta \\
g_{\beta \gamma} &= g_{\gamma \beta} = 0.
\end{align}
\end{subequations}
Here we introduced the dimensionless moments of inertia $\tilde{I}_i = I_i/I$. The metric (\ref{Orientation_state_metric}) leads to the metric determinant
\begin{equation}
g(\alpha, \beta, \gamma) = \sin^2 \beta,
\end{equation}
which is identical to the metric determinant (\ref{Metric_determinant_sphere}) arising for the motion on a sphere. From here on the phase-space representation for the orientation state is derived analogously to Sections \ref{Section:Curved_space_Wigner_function} and \ref{Section:Free_particle_on_a_sphere}, yielding the same expressions as in Ref.~\cite{Fischer2013a}.

\section{Conclusions} \label{Section:Conclusions}

We extended the Wigner-Weyl-Moyal phase-space representation to curved configuration spaces. All essential features of the standard formalism are maintained: The Wigner function can be interpreted as a quasiprobability distribution, and expectation values can be calculated by phase-space integrals, where Weyl-ordered operators are represented by their corresponding classical phase-space functions. Moreover, the quantum Liouville equation exhibits the correct semiclassical limit, which completes the desired connection with classical mechanics. Both unbounded and compact configuration spaces are covered; the latter display discrete momentum variables in phase space, which however exhibit a continuous semiclassical limit.

In contrast to previous approaches, we do not invoke possibly existing symmetries of the system, but consistently employ translations in order to construct the Stratonovich-Weyl operator kernels. This is at variance with Stratonovich's request for covariance \cite{Stratonovich1956a, Stratonovich1957a, Brif1999a}, expressing the expectation that the Weyl symbols are invariant under symmetry transformations. This is certainly a reasonable requirement in the Cartesian case, since the Wigner-Weyl-Moyal representation for a free particle with $\hat{H}=\hat{p}^2/2 m$ is invariant under translations. However, our results suggest that it is not the covariance which constitutes a defining feature of general phase-space representations, but rather that the use of the translation group is decisive to obtain a viable phase-space representation. The presented formalism can be applied to spaces of arbitrary curvature and thus opens the versatility of the quantum phase-space perspective to the wide field of curved configuration spaces.

{\it Acknowledgment:} T.F. and K.H. acknowledge funding from the DFG (HO2318/4).

%\bibliographystyle{h-physrev}
%\bibliography{literature}

\end{document}